\documentclass[a4paper,twocolumn,groupedaddress,showpacs]{revtex4}
\usepackage[]{graphicx}

\begin{document}


\title{Competition of ferromagnetism and superconductivity in Sc$_3$InB}

\author{B. Wiendlocha}
\email[email: ]{bartekw@fatcat.ftj.agh.edu.pl}
\thanks{\\ This work was presented at the European Conference {\it Physics of Magnetism'05}, 24-27.06.2005, Poznan, Poland.}
\author{J. Tobola}
\author{S. Kaprzyk}
\affiliation{Faculty of Physics and Applied Computer Science, 
AGH University of Science and Technology, al. Mickiewicza 30, 30-059 Krakow, Poland}
\author{D. Fruchart}
\affiliation{Laboratoire de Cristallographie, CNRS, BP 166, 38042 Grenoble Cedex 9, France}
\author{J. Marcus}

\affiliation{Laboratoire d'Etude des Proprietes Electroniques des Solides, CNRS, BP 166, 38042 Grenoble Cedex 9, France}
\begin{abstract}
We present results of electronic structure calculations for intermetallic 
perovskite Sc$_3$InB with 
Full--Potential KKR-LDA method. 
Sc$_3$InB is very promising candidate for a new superconductor (related 
to 8~K MgCNi$_3$) and can be regarded as a boron--inserted 
cubic Sc$_3$In, which is a high--pressure allotropic form of the 
hexagonal weak ferromagnet Sc$_3$In.
We predict that cubic Sc$_3$In can be also magnetic, whereas Sc$_3$InB 
having large DOS in the vicinity of E$_{F}$ exhibits non-magnetic ground state. 
Estimation of the electron--phonon coupling 
for Sc$_3$InB gives $\lambda \simeq 1$. Furthermore, the effect of vacancy 
in Sc$_3$InB$_{1-x}$ and antisite disorder in Sc$_3$(In-B) on critical 
parameters is also discussed in view of KKR--CPA method. 
All theoretical 
results support possibility of the 
superconductivity onset in Sc$_3$InB. 
Preliminary experimental measurements established 
the transition temperature close to 4.5~K, with a 
very abrupt change 
in susceptibility and a correlated drop of the 
resistivity when cooling down.
\end{abstract}
\maketitle                   

\section{Introduction}

Recent discovery of superconductivity in the intermetallic perovskite 
MgCNi$_3$~\cite{nature1} with T$_c \simeq 8$ K was a great surprise 
due to large Ni contents and this compound was rather expected to be 
near ferromagnetic critical point~\cite{pick1}. It was established that 
electron--phonon mechanism is responsible for superconductivity in 
this material, but few details are still not clear. NMR $t_1$ relaxation 
time experiments~\cite{nmr} or specific heat measurements~\cite{spheat} have 
resulted in typical behaviors, supporting {\it s}--wave type pairing 
with electron--phonon coupling constant $\lambda \sim 0.8$. 
Conversely, rather unconventional behaviors have been observed in other 
experiments (e.g. increase of critical temperature with pressure 
\cite{press1} or unusual low--temperature 
behavior of London penetration depth $\lambda_L(T)$~\cite{penetr}), which 
can be partly connected with spin fluctuations likely appearing due to 
vicinity of ferromagnetism. Moreover, complex dynamical properties of 
this superconductor, as e.g. soft--mode behaviors and instability of 
Ni vibrations~\cite{sav1},~\cite{ph}, make its theoretical analysis 
quite cumbersome.

In this paper we report on predictions of superconductivity in related 
Sc$_3$InB compound, which also seems to  be close to magnetism limit due 
to weak ferromagnetic properties of Sc$_3$In. It was already revealed 
that Sc$_3$InB crystallizes in a perovskite structure (space group 
$P$m-3m, CaTiO$_3$ type) with lattice constant $a = 4.56$ \AA~\cite{cryst}. 
However, the same number of valence electrons of boron and indium elements may give rise to 
lattice instabilities and In/B antisite defects are plausible.

\section{Theoretical study}

\begin{figure}[htb]
\includegraphics[width=0.35\textwidth]{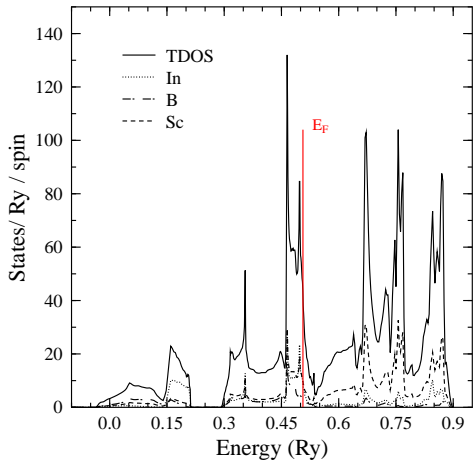}
\caption{Total electron DOS in Sc$_3$InB.}
\label{fig:1}
\end{figure}

Electronic structure calculations were performed by Full Potential KKR method 
\cite{kkr}, \cite{kkr2} within the LDA framework employing von Barth-Hedin formula for 
the exchange-correlation potential. Figure~\ref{fig:1} presents electron 
density of states (DOS) in Sc$_3$InB compound. The Fermi level (E$_{F}$) 
is located on the decreasing slope of large DOS peak with the $n(E_F)$ 
value being as large as $\simeq 90 $ states/Ry. The main contributions 
can be attributed to Sc ($d$--states) and In ($p$--states) 
(see Tab.~\ref{tab:1}). Noteworthy, appearance of a large DOS peak close 
to $E_{F}$, coming essentially from transition metal Ni--$d$ states, was also 
characteristic of electronic structure of MgCNi$_3$ (Mg contribution 
was negligible) \cite{pick1},\cite{szajek}. The Sc$_3$InB case is not similar, since 
In plays more active role in formation of electronic states near 
$E_{F}$ due to one electron more on $p$--shell.

\begin{table*}[htb]
\caption{Properties of Sc$_3$InB. $n(E_F)$ is given in Ry$^{-1}$/atom, $\eta$ in mRy/${a_B}^2$/atom,  $\omega$ in meV, $M\langle \omega^2 \rangle$ in mRy/${a_B}^2$.} \label{tab:1}
\begin{ruledtabular}
\begin{tabular}{lcccccccccccc}
Atom & $n(E_F)$ & $n_s(E_F)$ &$n_p(E_F)$ &$n_d(E_F)$ & $n_f(E_F)$ & $\eta$ & $\eta_{sp}$ & $\eta_{pd}$ & $\eta_{df}$ & $\sqrt{\langle \omega^2 \rangle}$ & $ M\langle \omega^2 \rangle $ & $\lambda$\\
\hline
Sc   &  20.26 & 0.06& 2.34 & 16.98& 0.88  & 18.6 &  0.0 &  4.7 & 13.3 & 18.5 & 75.7 & 0.74  \\
In   &  18.28 & 0.04& 16.72& 1.16 & 0.32  &  1.5 &  0.0 &  1.5 &  0.0 & 18.5 & 193.4 & 0.01  \\
B    &  6.72  & 0.02& 5.92 & 0.42 & 0.36  & 34.3 &  0.0 & 34.3 &  0.0 & 58.3 & 180.9 & 0.19  \\ 
\end{tabular}
\end{ruledtabular}
\end{table*}


\paragraph{Predictions of superconductivity in Sc$_3$InB}
The electron--phonon coupling strength was estimated by calculating
McMillan--Hopfield parameters~\cite{mcm} from Gaspari--Gy\"orffy 
formulas~\cite{gasgy} --~\cite{pick2} within Rigid Muffin Tin 
Approximation (RMTA). The electron--phonon coupling constant was 
then calculated from relation: $\lambda = \sum_i {\eta_i}/{m_i\langle \omega_i^2 \rangle}$, 
where $i$ corresponds to atoms in the unit cell. 
The $\langle \omega_i^2 \rangle$ parameter was derived from 
phonon DOS $F(\omega)$ \footnote{definition: $\langle 
\omega^n \rangle = \int \omega^{n-1} \alpha^2 F(\omega) d\omega / 
\int \omega^{-1} \alpha^2 F(\omega) d\omega \simeq \int 
\omega^{n-1} F(\omega) d\omega / \int \omega^{-1} F(\omega) d\omega$, 
if $\alpha^2(\omega) \simeq const.$, where $\alpha^2(\omega)$ is the electron--phonon interaction coefficient.}, computed for minimum--energy lattice constant $a_0 = 8.610$ $a_B$ (1~$a_B$ = 0.529 \AA) within Density Functional Perturbation Theory, using the PWscf package~\cite{pwscf}.

In order to take into account the markedly different masses of Sc, In and B atoms, phonon spectrum was analyzed 
while focusing on two parts: low frequency region with predominantly In and Sc modes 
(peak at 13~meV comes from flat acoustic In modes) and high frequency region 
with B vibrations (above 50~meV). Noteworthy, similar separation of 
phonon DOS was earlier observed experimentally in MgCNi$_3$~\cite{ph}. 
Values of $\langle \omega_i^2 \rangle$ used next to estimate $\lambda$, were calculated separately for both regions (with cut line at 50~meV): $\langle \omega_i^2 \rangle$ for Sc and In are taken to be equal, and represent the low--frequency part, while $\langle \omega_B^2 \rangle$  was taken from high frequency part of the spectrum. Results and estimation of $\lambda_i$ are presented 
in Table~\ref{tab:1}, total electron--phonon coupling constant 
is $\lambda = 0.94$.

Using McMillan formula~\cite{mcm} for critical temperature and applying 
typical value of Coulomb pseudopotential $\mu^{\star} = 0.13$, we got 
$T_c \simeq 12$ K (we used $\langle \omega \rangle/1.2$ with $\langle \omega \rangle = 19.5$~meV 
instead of $\omega_D/1.45$ in McMillan formula in practical computations). 
Both $\lambda$ and $T_c$ values belong to moderate regime 
of superconducting parameters within the RMTA model, and are higher than in~MgCNi$_3$.

\begin{figure}[htb]
\includegraphics[width=0.35\textwidth]{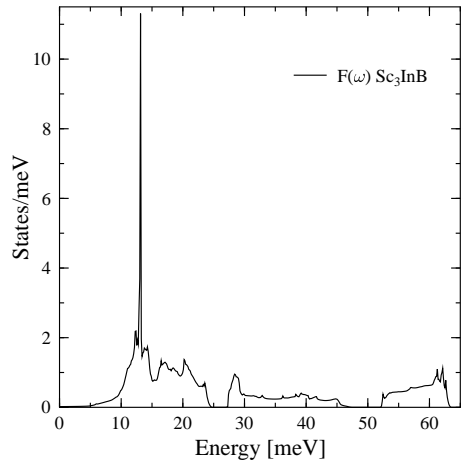}
\caption{Phonon DOS in Sc$_3$InB.}
\label{fig:2}
\end{figure}

We can also notice advantageous trend of electron--phonon coupling 
in Sc$_3$InB if comparing with MgCNi$_3$ superconductor. 
Our KKR calculations for MgCNi$_3$ showed that Ni has the largest 
McMillan--Hopfield parameter: $\eta_{Ni} = 20$ mRy/$a_B^2$ with respect to
other atoms ($\eta_{C} = 9$ mRy/$a_B^2$, $\eta_{Mg}$ - negligible). 
The value of $\eta_{Sc}$ in Sc$_3$InB is similar to $\eta_{Ni}$ 
in MgCNi$_3$ but $\eta_{B}$ is over three times larger than 
$\eta_{C}$ (both In and Mg contributions are much smaller).
In view of recent C isotope effect measurements~\cite{isef} (very large 
$\alpha_C = 0.54$ coefficient) and bearing in mind a particular sensitivity 
of $T_c$ on the carbon concentration, one can conclude that both C and 
Ni sublattices are important in superconductivity of MgCNi$_3$. 
In view of these criterion $T_c$ in Sc$_3$InB can be expected 
higher than in~MgCNi$_3$.

\paragraph{Magnetic properties of Sc$_3$In}
It seems that instability towards magnetism may also be present in 
Sc$_3$InB$_{1-x}$ compound. First, we have studied electronic structure 
of both allotropic phases of Sc$_3$In. The hexagonal compound 
(Ni$_3$Sn-type, $a$ = 6.42~\AA, $c$ = 5.18~\AA \ \cite{cryst}) is well--known 
weak itinerant ferromagnet, while the cubic compound 
(Cu$_3$Au--type, $a = 4.46$ \AA, synthesized under high pressure \cite{cryst}) has 
not been yet investigated to our knowledge. Present KKR calculations 
showed that both allotropic forms of Sc$_3$In should exhibit magnetic 
ground state supported by magnetic moment on Sc atoms, i.e. 0.26 $\mu_B$ 
(hexagonal phase) and 0.27 $\mu_B$ (cubic phase). 
We should remind that experimentally observed magnetic moment 
is much weaker ($\sim 0.05 \mu_B$/Sc in hexagonal phase), as already 
underlined in the previous LAPW calculations~\cite{singh}. 
Besides, the superconductor--to--ferromagnet transition can appear in 
Sc$_3$InB$_{1-x}$ if varying B content (KKR-CPA computations are in progress). Thus, as already suggested 
in MgCNi$_3$~\cite{pick1}, \cite{sav1} one can expect that the proximity of 
ferromagnetic quantum critical point (resulting in enhanced spin 
fluctuations) may probably compete with superconductivity. 

\section{Experimental analyses}

A~$\sim$1.5~g sample was first prepared by arc melting under high purity argon atmosphere (99.9995) starting from the appropriate proportions of the elements (purity $>$~99.95) to obtain the Sc$_3$InB formula. The resulting small ingot was melted several times in order to insure homogeneity. Pieces of the ingots were made by using a metal mortar, and inspection by using an optical microscope reveals the bright and homogeneous aspect of the fractured surfaces. Then, XRD patterns were recorded at $\lambda_{K\alpha}(Cu)$ using a Bragg-Brentano diffractometer equipped with a backscattering pyrolitic graphite monochromator. The diffraction pattern reveals the presence of a dominant amount of cubic phase with the addition of a minor impurity. Probably because no annealing procedure was applied, the crystallised state of the sample was not of the best and no effective crystal structure refinement was applied. However using the PowderCell code~\cite{pcell}, the two main phases were clearly identified and a rough determination of the cell parameters was made possible. The main phase ($\sim$70~\% vol.) is simple cubic of perovskite type of structure and the second one ($\sim$30~\% vol.)  is hexagonal of Ni$_2$In type of structure (space group $P6_3/mmc$). Accounting for the large difference in between the scattering lengths of the p-elements B and In, an estimate of the composition of the main phase was made. All the results are displayed on Table~\ref{tab:cryst}.

\begin{table}[htb]
\caption{Structure analysis of the material synthesized with the nominal composition Sc$_3$InB.} \label{tab:cryst}
\begin{ruledtabular}
\begin{tabular}{cccccc} 
Compound & Type & \multicolumn{3}{c}{Cell parameters [\AA]} & Composition\\ \hline
Sc$_3$InB \\($\sim 70$\% vol.) &  CaTiO$_3$ & 4.66 & 4.66 & 4.66& Sc$_3$In$_{1.3}$B$_{0.7}$ \\ \hline
Sc$_2$In \\ ($\sim 30$\% vol.) & Ni$_2$In & 5.05 & 5.05 & 6.30 & Sc$_2$In\\ 
\end{tabular}
\end{ruledtabular}
\end{table}

The best agreement for the main phase composition leads to consider that the p--elements B and In are neither fully ordered nor randomly distributed, with about 0.70~In and 0.30~B atom on the 1a site, then 0.65~In and 0.35~B atom on the 1b site of the space group $P$m-3m. Consideration for such a type of atomic disordering was already reported in literature~\cite{d2}. Both the compositions of the two compounds (3-1-1~and~2-1) confirm that boron is a difficult element to combine, thus the remaining boron should be detected in the XRD pattern, unless it is difficult to evidence effectively as a very light and often poorly crystallised element.

\begin{figure}[htb]
\caption{A.c. susceptibility trace revealing the transition to superconducting state.}
\includegraphics[width=0.35\textwidth]{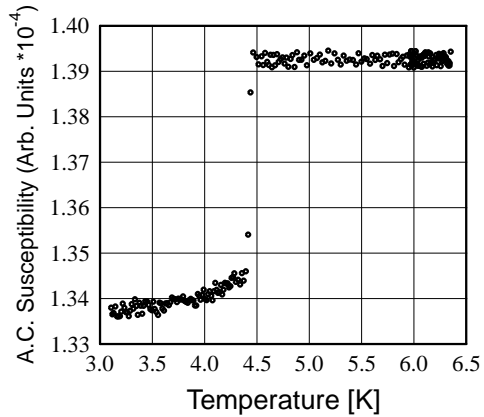}
\label{fig:exp}
\end{figure}

Susceptibility and resistivity measurements were made using a a.c. susceptometer in temperature ranging from 50 to 2~K. Millimeter sized pieces of the ingot were measured successively and lead to the same results. A typical record is shown on Figure~\ref{fig:exp}, thus revealing the onset of a superconducting state down to 4.4~K, the transition being very sharp with no detectable hysteresis loop. A change in the resistivity trace was also observed simultaneously.

Unexpectedly, the rest of the ingot and the small pieces rapidly change their brittle and bright aspects after several hours let in ambient atmosphere. Besides, a tentative to melt twice the main parts of the initial ingot was made, but at this time no evidence for any superconducting transition was found. Then, several new syntheses were undertaken, thus operating as possible similarly as for the first one procedure. Again, the new ingots we obtained do not display any transition down to 2~K. As revealed by the theoretical derivations (see below), the superconducting state and the related transition look fairly dependant on the p--element ordering and the stoichiometry in the 1a and 1b sites. So, we anticipate that during and after the different melts, the new samples do not exhibit the same atom ordering as it was resulting from the first attempt. New syntheses are now scheduled to be undertaken using different methods and techniques in order to achieve optimized compounds.

Because of experimental problems with synthesis,
effect of boron site vacancy on electronic structure near 
E$_{F}$ and superconducting properties was simulated using 
KKR method with coherent potential approximation (CPA). 
We have found significant change in $\eta$ values not only for B, 
but also for Sc, i.e. for 7\% boron deficiency (Sc$_3$InB$_{0.93}$), 
employing the same values of $\langle \omega_i^2 \rangle$, 
total $\lambda$ decreased over 30\% to 0.62, which resulted in $T_c \simeq 4$ K. 
In Sc$_3$InB$_{0.85}$ coupling constant is so small ($\lambda \sim 0.4$), 
$T_c < 0.5$~K seems to be below the standard low--temperature 
measurements. 
Similar decrease (but less rapid than in the case of B vacancy) 
of critical parameters was detected from KKR-CPA analysis when 
antisite In/B disorder increased.

\end{document}